\documentclass[aps,prl,twocolumn,superscriptaddress]{revtex4-2}

\usepackage{amsmath}
\usepackage{amsfonts}
\usepackage{amssymb}
\usepackage{graphicx}
\usepackage{bm}
\usepackage{color}
\usepackage{multirow}
\usepackage{lineno}
\usepackage{dsfont}
\usepackage{siunitx} 
\usepackage{lipsum}
\usepackage{array}
\usepackage{newtxtext,newtxmath}

\usepackage{verbatim}
\usepackage{mathtools}
\usepackage{empheq}

\usepackage[normalem]{ulem}  
\usepackage{cancel}

\definecolor{darkgreen}{rgb}{0.09, 0.55, 0.3}

\definecolor{darkred}{rgb}{0.8, 0.10, 0.1}

\newcommand{\bea}{\begin{eqnarray}}
\newcommand{\eea}{\end{eqnarray}}
\newcommand{\be}{\begin{equation}}
\newcommand{\ee}{\end{equation}}
\newcommand{\bes}{\begin{equation*}}
\newcommand{\ees}{\end{equation*}}
\newcommand{\bi}{\begin{itemize}}
\newcommand{\ei}{\end{itemize}}

\renewcommand{\vec}{\mathbf}
\DeclareSymbolFont{usualmathcal}{OMS}{cmsy}{m}{n}
\DeclareSymbolFontAlphabet{\mathcal}{usualmathcal}
\newcommand{\rr}{\vec{r}}

\newcommand{\kb}{\ensuremath{\text{k}_\mathrm{B}}}

\newcolumntype{P}[1]{>{\centering\arraybackslash}p{#1}}
\newcommand{\pr}[1]{\ensuremath{\left[#1\right]}} 
\newcommand{\pc}[1]{\ensuremath{\left(#1\right)}}

\newcommand{\ket}[1]{\ensuremath{\left\vert#1\right\rangle}} 
\newcommand{\md}[1]{\ensuremath{\left\vert#1\right\vert}}

\definecolor{bluedarkRL}{rgb}{0,0.3,0.79}

\renewcommand{\Re}{\operatorname{Re}}

\newcommand{\lvdW}{\ensuremath{\ell_\mathrm{vdW}}}
\newcommand{\ldb}{\ensuremath{\lambda_\mathrm{T}}}
\newcommand{\Gb}{\ensuremath{\Gamma_\mathrm{b}}}

\definecolor{burgundy}{rgb}{0.5, 0.0, 0.13}
\definecolor{denim}{rgb}{0.08, 0.38, 0.74}
\definecolor{midnightgreen}{rgb}{0.0, 0.29, 0.33}
\definecolor{sienna}{rgb}{0.53, 0.18, 0.09}
\definecolor{sacramentostategreen}{rgb}{0.0, 0.34, 0.25}

\newcommand{\mi}{\mathrm{i}}

\DeclareSIUnit\gauss{G}
\definecolor{Green}{rgb}{0,0.6,0.4}

\usepackage{xcolor}
\usepackage[%
  colorlinks=true,
  urlcolor=bluedarkRL,
  linkcolor=bluedarkRL,
  citecolor=bluedarkRL
]{hyperref}
\hypersetup{
colorlinks=true,
linkcolor=bluedarkRL,
citecolor=bluedarkRL,
urlcolor=black}

\begin{document}

\title{Two-Body Contact Dynamics in a Bose Gas near a Fano-Feshbach Resonance}

\author{Alexandre Journeaux}
\affiliation{Laboratoire Kastler Brossel, Coll\`ege de France, CNRS, ENS-Universit\'e PSL, Sorbonne Universit\'e, 11 Place Marcelin Berthelot, 75005 Paris, France}
\author{Julie Veschambre}
\affiliation{Laboratoire Kastler Brossel, Coll\`ege de France, CNRS, ENS-Universit\'e PSL, Sorbonne Universit\'e, 11 Place Marcelin Berthelot, 75005 Paris, France}
\author{Maxime Lecomte}
\affiliation{Laboratoire Kastler Brossel, Coll\`ege de France, CNRS, ENS-Universit\'e PSL, Sorbonne Universit\'e, 11 Place Marcelin Berthelot, 75005 Paris, France}
\author{Ethan Uzan}
\affiliation{Laboratoire Kastler Brossel, Coll\`ege de France, CNRS, ENS-Universit\'e PSL, Sorbonne Universit\'e, 11 Place Marcelin Berthelot, 75005 Paris, France}
\author{Jean Dalibard}
\affiliation{Laboratoire Kastler Brossel, Coll\`ege de France, CNRS, ENS-Universit\'e PSL, Sorbonne Universit\'e, 11 Place Marcelin Berthelot, 75005 Paris, France}
\author{Félix Werner}
\affiliation{Laboratoire Kastler Brossel, Coll\`ege de France, CNRS, ENS-Universit\'e PSL, Sorbonne Universit\'e, 11 Place Marcelin Berthelot, 75005 Paris, France}
\author{Dmitry S. Petrov}
\affiliation{Universit\'e Paris-Saclay, CNRS, LPTMS, 91405, Orsay, France}
\author{Raphael Lopes}
\email{raphael.lopes@lkb.ens.fr}
\affiliation{Laboratoire Kastler Brossel, Coll\`ege de France, CNRS, ENS-Universit\'e PSL, Sorbonne Universit\'e, 11 Place Marcelin Berthelot, 75005 Paris, France}

\begin{abstract}
  We investigate the real-time buildup of short-range correlations in a nondegenerate ultracold Bose gas near a narrow Fano-Feshbach resonance. Using rapid optical control, we quench the closed-channel molecular energy to resonance on submicrosecond timescales and track the evolution of the two-body contact through photodissociation losses. Repeated pulse sequences enhance sensitivity to early-time two-body losses and reveal long-lived coherence between atom pairs and molecular states. The observed dynamics are accurately reproduced by our dynamical two-channel zero-range theory, which explicitly accounts for the resonance’s narrow width and finite closed-channel decay, establishing a predictive framework for correlation dynamics in quantum gases near Fano-Feshbach resonances.
 \end{abstract}
\maketitle




Understanding and predicting out-of-equilibrium dynamics in quantum many-body systems is one of the central challenges of modern physics. Ultracold gases offer a pristine test bed to address this challenge, with well-defined initial states, tunable interactions, and microscopic dynamics that can be resolved directly in experiment. This unique level of control has enabled landmark observations of nonequilibrium phenomena in quantum gases~\cite{Greiner2002, Cetina2016,Bernien2017,Prufer2018,Erne2018,Senaratne2018,Skou2021,Etrych2025,Vivanco2025}.

Among the various nonequilibrium settings explored so far, a paradigmatic case  is the three-dimensional Bose gas quenched from the weakly interacting regime to the maximally interacting unitary limit, whose experimental study~\cite{Makotyn2014,Fletcher2017,Eigen2017,Klauss2017,Eigen2018,Zhang2023} stimulated extensive theoretical activity~\cite{Natu2013,Sykes2014,Corson2015,Dincao2018,Munoz2019,Colussi2020,Sun2020,Gao2020,Qi2021,Ahmed-Braun2022,Yang2023,Cui2024,VandeKraats2024,Wang2024,Wang2025}. The dynamics of this system are governed by the buildup of short-range two-body correlations, which are naturally described in terms of Tan’s contact $\mathcal{C}_2$~\cite{Tan2008Momentum,Tan2008,Tan2008Generalized,Braaten2008,Zhang2009,Braaten2011bosons, Werner2012bosons,Braaten2012}, a unifying concept in quantum gases with close connections to nuclear physics~\cite{BarneaNuclContactsPRL2015,HenPLB2018,Nucl_C_exp_Nature2020,HenContactsNatPhys2021}. The contact $\mathcal{C}_2$, proportional to the probability of finding two atoms at short interparticle distances, connects microscopic and thermodynamic quantities through exact relations.
While $\mathcal{C}_2$ has been measured at equilibrium~\cite{Wild2012,Fletcher2017,Zou2021}, its nonequilibrium dynamics remain elusive~\cite{Makotyn2014,Eigen2018}, most likely because of additional three-body contributions~\cite{Smith2014,VandeKraats2024}.

A natural probe of $\mathcal{C}_2$ is provided by microscopic loss mechanisms, such as photodissociation of closed-channel molecules, which have long served as sensitive diagnostics of short-range correlations at equilibrium~\cite{Partridge2005,Werner2009,Wang2021,Jager2024}. These processes directly reflect the amplitude of short-range correlations while minimally perturbing the many-body state, making them a promising route to probe correlation dynamics in real time. However, exploiting them to follow the growth of correlations requires interaction control on timescales shorter than the buildup time, which remains a significant experimental challenge.

Probing the dynamics of $\mathcal{C}_2$  can be carried out near both broad and narrow Fano-Feshbach resonances (FFRs). While broad resonances are fully characterized by the scattering length~\cite{ChinRMP10}, narrow resonances enrich this description by introducing an additional length scale $R^\star$ set by the effective range and inversely proportional to the resonance width~\cite{Petrov2004,Braaten2008PRA,Pricoupenko2013}. This distinctive feature makes them a powerful platform to explore effective-range effects, which are of central importance in low-energy nuclear systems~\cite{Schwenk2005,Gezerlis2012,Platter2025}.




In this Letter, we implement a new approach based on optical tuning of the closed-channel molecular state, enabling genuine sudden quenches on submicrosecond timescales and providing access to previously unresolved contact dynamics. This fast control is achieved via spin-dependent light shifts arising from the large vectorial and tensorial polarizability of lanthanide atoms \cite{Becher2018,Chalopin2018,Kreyer2021,Note1}\footnotetext{A similar mechanism has been used in a Li-K mixture \cite{Cetina2016}. In that case, the light shift also acts in the open channel, and special care is required to avoid modifications of the trapping frequencies.}. 
By periodically cycling between resonant and weak interactions, we enhance sensitivity to early-time two-body losses and resolve the buildup of short-range correlations following an interaction quench.
We develop a two-channel model that incorporates the narrow resonance width and molecular-state dissipation, relates the two-body loss rate to the instantaneous contact $\mathcal{C}_2(t)$, and accurately reproduces the observed dynamics.
Finally, by varying the off-resonant time, we observe oscillations of the two-body loss rate, extending previous observations of interference effects arising from the coherent superposition of dimer and unbound atomic states~\cite{Donley2002,Claussen2003, Elbaz2023}.





  \begin{figure}[t!]
    \centering
	\includegraphics[width=\columnwidth]{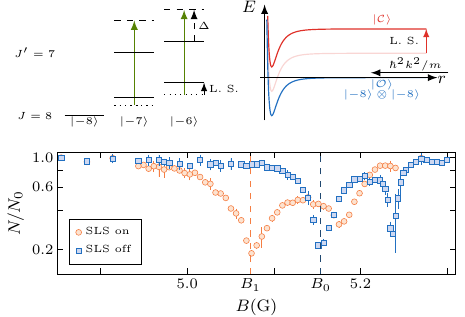}
	\caption{Fano-Feshbach resonance (FFR) displacement induced by a spin-dependent light shift. Top: illustration of the coupling between the ground-state manifold with total angular momentum $J=8$ and the excited-state manifold associated with the transition at wavelength $\lambda = \SI{530.306}{\nano\meter}$, with $J' = 7$. This coupling induces a light shift that displaces the energy of the closed-channel molecular state $\ket{\mathcal{C}}$, while the open-channel collisional state $\ket{\mathcal{O}}$ remains unaffected. Bottom: loss features in the presence (disks) and absence (squares) of the spin-dependent light shift (SLS) laser, showing the optical displacement of the magnetic Fano-Feshbach resonance from $B_0$ to $B_1$ for a thermal sample of $^{162}$Dy with $N=3\times 10^4$ atoms and temperature $T= \SI{1.24}{\micro\kelvin}$. In the ``SLS on'' (resp. ``SLS off'') case we hold the sample for \SI{15}{\milli\second} (resp. \SI{150}{\milli\second}).}
    \label{figscheme}
\end{figure}

We perform the experiment with $^{162}$Dy atoms and use a laser beam detuned by $\Delta = 2\pi \times \SI{20}{\giga\hertz}$ from the optical transition at $\lambda = \SI{530.306}{\nano\meter}$, which couples the ground-state manifold with total angular momentum $J=8$ to an excited state with $J' = J - 1$~\cite{Lecomte2025} (see Fig.~\ref{figscheme}). This  beam is linearly polarized along $\hat{z}$, resulting in a light shift for most closed-channel molecular states \cite{Bauer2009Exp, Jagannathan2016}. By contrast, the open-channel state, composed of atoms in the spin configuration $\ket{-8} \otimes \ket{-8}$, remains effectively decoupled from the light field. As a result, we do not observe heating or atom loss due to spontaneous emission when operating far from the FFR. 

This spin-dependent light shift (SLS) leads to a displacement of the FFR pole from its original position $B_0$ to a new magnetic field $B_1$. The shift is given by $B_1 - B_0 \propto - {I}/(\Delta \delta \mu)$, where $I$ is the laser intensity and $\delta\mu>0$ the differential magnetic moment between the molecular and atomic states.

Our experiments start with a thermal sample of typically $10^5$ atoms. As shown in Fig.~\ref{figscheme}, the SLS beam shifts the FFR loss feature from $B_0 \approx \SI{5.15}{\gauss}$ to $B_1 \approx \SI{5.07}{\gauss}$. This shift enables a rapid change in scattering length, from $a\approx \SI{140}{}\,a_0$ to resonance ($a \to \infty$), within \SI{200}{\nano\second} \cite{Cetina2016,Note2}\footnotetext{For comparison, this is an order of magnitude faster than the state-of-the-art magnetic field quench time for broad FFRs~\cite{Kell2021}.}.
In addition to shifting the resonance, the laser also induces two-body losses due to photodissociation of closed-channel molecules, which we use here as a tool for probing short-range correlations.
The laser-induced energy shift of the closed-channel molecule and the associated two-body loss thus provide a method to probe the dynamics of short-range two-body correlations.



\begin{figure}[t!]
    \centering
	\includegraphics[width=\columnwidth]{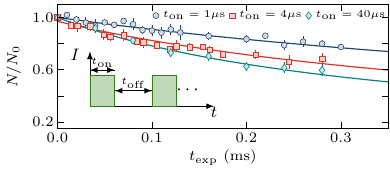}
	\caption{Probing two-body loss dynamics. Atom number as a function of exposure time, $t_\mathrm{exp} = t_\mathrm{on} N_\mathrm{cycles}$, for three different values of $t_\mathrm{on}$ (see legend), at a temperature $T = \SI{0.34}{\micro\kelvin}$, and initial atom number $N_0 \approx 2 \times 10^4$. Here, $t_\mathrm{off} = \SI{26}{\micro\second}$. Solid lines are fits using Eq.~\eqref{eqtwobodyloss}. Inset: schematic of the pulsing protocol. }
    \label{lossmagnifier}
\end{figure}

To better resolve the dynamics we employ a repeated pulsing protocol that improves the detectability of two-body losses by accumulating their effects over many interaction quenches near resonance (see Fig.~\ref{lossmagnifier} inset). This approach enhances sensitivity to small loss rates without modifying the underlying inelastic mechanism.
Specifically, we prepare a thermal gas at the magnetic field $B_1$, and apply a sequence of optical pulses: the SLS beam is turned on for a duration $t_\mathrm{on}$, followed by an off period $t_\mathrm{off}$ long enough to ensure that successive pulses act independently. This sequence is repeated $N_\mathrm{cycles}$ times, resulting in a total exposure time $t_\mathrm{exp} = t_\mathrm{on}  N_\mathrm{cycles} $ \cite{SM} \nocite{Rem2013,Pricoupenko2019}. 

In Fig.~\ref{lossmagnifier}, we show the atom number evolution as a function of $t_\mathrm{exp}$ for a thermal sample with temperature $T = \SI{0.34(1)}{\micro\kelvin}$ and three values of $t_\mathrm{on}$ with $t_\mathrm{off} = \SI{26}{\micro\second}$. For a fixed  $t_\mathrm{exp}$, we observe a monotonic increase in atom loss with increasing $t_\mathrm{on}$, a signature that the two-body loss rate is dynamically evolving. We fit the data using a two-body loss model,
\be
\frac{dN}{d t_\mathrm{exp}}
= -L_2 (t_\mathrm{on}, {t_\mathrm{off}}) \ \bar n \ N,
\label{eqtwobodyloss}
\ee
where $\bar n = n_0 / \sqrt{8}$ is the spatially averaged density, assuming a Gaussian spatial distribution with peak value $n_0$. We extract the effective two-body loss coefficient $L_2(t_\mathrm{on}, {t_\mathrm{off}})$, defined as
\be
L_2(t_\mathrm{on}, t_\mathrm{off}) = 
\frac{1}{t_\mathrm{exp}}
\int_0^{(t_\mathrm{on}+t_\mathrm{off})N_\mathrm{cycles}} \mathcal{L}_2(t) \ dt 
\label{eq:L2_vs_mathcal_L2}
\ee
where $\mathcal{L}_2$ is the instantaneous loss rate such that locally $\dot{n}=-\mathcal{L}_2\,n^2$.
The right-hand side of Eq.(\ref{eq:L2_vs_mathcal_L2}) would be independent of both $t_{\rm on}$ and $t_{\rm off}$ if $\mathcal{L}_2$ was constant during $t_{\rm on}$ and zero during $t_{\rm off}$. Accordingly, $L_2$ allows us to quantify deviations from this simple scenario.
Since atom loss is accompanied by heating, we restrict our fit to data where the temperature increase is below 15\% and take into account the small resulting change in effective volume due to thermal expansion of the cloud~\cite{SM}.

\begin{figure}[t!]
    \centering
	\includegraphics[width=\columnwidth]{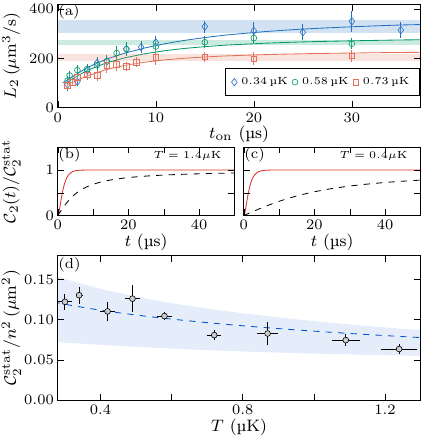}
	\caption{Two-body contact dynamics.
(a)~Measured two-body loss coefficient $L_2$ for different temperatures (see legends). Solid lines are fits using Eq.~\eqref{eqForphi3}. The shaded region indicates the value of $\bar L_2$ plus or minus one standard deviation.
(b) Evolution of $\mathcal{C}_2 (t)$, scaled by its stationary value, for $R^\star \to 0$ and $\Gb =0$ (black dashed line), $R^\star \to \infty$  and  $\Gb / 2\pi = \SI{300}{\kilo\hertz}$ (red line), for a thermal sample at \SI{1.4}{\micro\kelvin}. (c) Same comparison for a thermal sample at \SI{0.4}{\micro\kelvin}.
(d)~Temperature dependence of $\mathcal{C}_2^{{\rm stat}} / n^2$, deduced from $\bar{L}_2$, and compared with the theoretical prediction from Eq.~\eqref{eqC2narrowFFR}. 
The blue dashed line shows the prediction from Eq.~(\ref{eqC2narrowFFR}), with the values of $R^\star$ and $\Gb^\mathrm{on}$ given by Eq.~(\ref{values_Gb_Rs}) and the corresponding error bars reflected by the shaded area.} 
    \label{L2vsT_pole_comparison}
\end{figure}

As shown in Fig.~\ref{L2vsT_pole_comparison}(a), for $t_\mathrm{off} = \SI{26}{\micro\second}$, $L_2$ increases with $t_\mathrm{on}$ at short times and saturates for $t_\mathrm{on} \gtrsim \SI{20}{\micro\second}$ to an asymptotic value $L_2^\mathrm{stat}$. We independently confirm this saturation behavior using a continuous-probe configuration, where $\bar L_2 = L_2(t_\mathrm{off} = 0)$, shown as the shaded region in Fig.~\ref{L2vsT_pole_comparison}(a). The measured value of $\bar L_2$ agrees with $L_2^\mathrm{stat}$ within experimental uncertainty, reinforcing our interpretation that the pulsed sequence faithfully captures the buildup dynamics of two-body losses.




To model the observed dynamics, we reduce the many-body problem to a two-body problem, which can be justified in the nondegenerate regime by a virial expansion~\cite{Sun2020}, and we adopt a zero-range two-channel description.  In this framework, the wave function of the relative motion in the open channel $\Psi({\bf r},t)$ is coupled to the closed-channel amplitude $\phi(t)$, with coupling strength characterized by the range parameter $R^\star > 0$. 
We find that $\phi$ is related to the $1/r$ singularity of $\Psi$  by the equation $\phi (t)= \sqrt{4\pi R^\star} \lim_{r \rightarrow 0} r \,\Psi({\bf r},t)$.
As a consequence, the two-body contact is proportional to the density $n_b$ of closed-channel molecules, $\mathcal{C}_2(t) = (8\pi/R^\star)\,n_b(t)$, a result previously established at equilibrium~\cite{Werner2009,Castin2012,Note3}\footnotetext{There is a factor 2 difference compared to the fermionic case considered in Refs.~\cite{Werner2009,Castin2012},  because for fermions, $\mathcal{C}_2$ is defined as 
the prefactor of the $1/k^4$ tail of the {\it spin resolved} momentum distribution}.
Since $n_b(t)$ equals $n^2/2$ times the thermal average $\langle\md{\phi(t)}^2\rangle_T$, and with each molecule being lost at a rate $\Gb(t)$, the instantaneous atom loss rate constant is~\footnote{
Here we assume that $n$ is time independent as it enters in our theoretical analysis only through the density of pairs $n^2/2$ and evolves on a much longer timescale than the buildup dynamics we are interested in.}
\be
\mathcal{L}_2 (t) = \Gb(t) \ \frac{R^\star}{4\pi} \ \frac{{\mathcal{C}_2(t)}}{n^2} = \Gb(t) \, \langle |\phi(t)|^2 \rangle_T.
\label{eqL2}
\ee

The problem of calculating the loss rate thus reduces to determining $\phi(t)$ for a pair of particles in a unit volume. We find (see Ref.~\cite{SM}) that the evolution of $\phi(t)$ for two identical bosons with nonzero relative momentum $k$ is governed by
\begin{equation}
\label{eqForphi3}
i \hbar \dot{\phi} (t) -E_0(t) \phi(t)-\frac{\hbar^{3/2}}{\sqrt{i \pi m} \, R^\star} \hspace{-0.1cm}\int_{-\infty}^t \hspace{-0.1cm}\frac{\dot{\phi}(t') \mathrm{d}t'}{\sqrt{ t-t'}} = -\frac{\hbar^2}{m} \sqrt{\frac{8\pi}{R^\star}} e^{-i \frac{\hbar k^2}{m} t} ,
\end{equation}
where $E_0(t) = -\hbar^2/\pr{m\, a'(t) R^\star} - i \hbar \Gb(t)/2$ is the complex closed-channel molecular energy detuning, and $1/a'(t)$ is the real part of the inverse scattering length. 

In the stationary regime, substituting $\phi(t)=\phi^{\rm stat} e^{-i \hbar k^2 t/m}$ into Eq.~(\ref{eqForphi3}) yields an algebraic equation, whose solution gives the stationary contact density 
\be
\mathcal{C}_2^{\rm stat} = n^2 \, \ldb^2 \int_0^\infty \mathrm{d}x \ \frac{32 \sqrt{2}\, x^2 e^{-x^2/(2\pi)}}{( \ldb/a' + x^2 R^\star/\ldb)^2 + ({\frac{m}{2\hbar}}\Gb \ldb R^\star + x)^2},
\label{eqC2narrowFFR}
\ee
where $\ldb = \sqrt{2\pi\hbar^2/ (m \kb T)}$ is the thermal de Broglie wavelength. Equation~\eqref{eqC2narrowFFR} shows that $\mathcal{C}_2^{\mathrm{stat}}$ depends on three dimensionless parameters characterizing, respectively, the dimer detuning, the resonance width, and the loss strength. 
For a broad lossless resonance  ($R^\star \to 0,\, \Gb \to 0$), $\mathcal{C}_2^{{\mathrm{stat}}}$ becomes independent of the sign of $a'$ and peaks at the resonance position, where we recover the universal result $\mathcal{C}_2^{{\mathrm{stat}}} = 32\pi n^2 \ldb^2$ \cite{Smith2014}. By contrast, near a narrow resonance, the maximum of $\mathcal{C}_2^{\rm stat}$ (and hence of $L_2^{{\rm stat}}$) shifts to a finite negative scattering length ($\ldb/a' < 0$)~\cite{SM}.

When $a'$ and $\Gb$ are suddenly quenched, the field $\phi$ evolves toward a new stationary state. As one can see from Eq.~\eqref{eqForphi3}, the dynamics and associated timescales depend on several parameters. For a single quench at $t=0$ from a very large initial detuning to a value $E_0$ close to resonance we can neglect $\phi(t)$ at $t<0$ and solve Eq.~(\ref{eqForphi3}) analytically by the Laplace transform technique. 
One can check that the transform of $\phi(t)$ explicitly reads 
\be
\phi(s)=\frac{\sqrt{8\pi/R^\star}}{{(k^2-ims/\hbar)\, (s+iE_0/\hbar+\sqrt{i\hbar s/m}/R^\star)}}
\label{eqLaplace}\, .
\ee 
The corresponding inverse Laplace transform is in general a bulky but analytic expression written in terms of error functions. It depends on the dimensionless parameters $m E_0/\hbar^2 k^2$ (which is complex) and $k R^\star$, and we can choose, for example, $m/\hbar k^2$ as the overall timescale. For thermally averaged quantities we use the same parameter space with $m/\hbar k^2$ replaced by $\tau_{\rm T}=\hbar/\kb T$.
Let us discuss the dynamics after a quench to resonance, i.e., $\Re(E_0) = 0$, in two opposite limits described by compact formulas.




In the lossless ($\Gb \ll 1/\tau_{\rm T}$), broad-resonance ($R^\star \ll \lambda_{\rm T}$) limit, we can neglect the first two terms in Eq.~(\ref{eqForphi3}), or the corresponding terms in Eq.~\eqref{eqLaplace}, and arrive at $(4\pi/R^\star)\md{\phi(t)}^2=32\,\pi^2\,\big|{\rm Erf}\sqrt{-i{\hbar k^2t/m}}\,\big|^2/k^2$, where ${\rm Erf}$ is the error function. This quantity oscillates in time, but averaging over momenta leads to $\mathcal{C}_2(t)~=~64\,n^2\,\ldb^2\,{\rm arctan}({t/\tau_T})$~\cite{Cui2024}.

As a second example, we consider the lossy ($\Gb \gg 1/\tau_{\rm T}$), narrow-resonance limit ($R^\star \gg \lambda_{\rm T}$), where we can neglect the integral term in Eq.~(\ref{eqForphi3}). Then, 
$ \phi(t) = \mi \sqrt{8\pi/ R^\star} \ (e^{-i{\hbar k^2t/m}}-e^{-\Gb t/2})/(m \Gb/2\hbar - ik^2)$ and, after thermal averaging, we obtain
$\mathcal{C}_2(t)=(1-e^{-\Gb t/2})^2 \ \mathcal{C}_2^{{\mathrm{stat}}}$.
We note that, in this case, $\mathcal{C}_2(t)\propto t^2$ at short times~\cite{Ahmed-Braun2022}, different from the $\mathcal{C}_2(t)\propto t$ scaling in the broad-resonance case~\cite{Sykes2014,Corson2015,Qi2021,Cui2024}. In Figs.~\ref{L2vsT_pole_comparison}(b)-(c) we present a few examples of the time dependence of the contact density predicted for various experimental parameters following a single quench to resonance.

Using the periodic pulse protocol, we now experimentally and theoretically reveal the dynamics of the two-body contact.
We examine the temperature dependence of the two-body loss rate dynamics by probing thermal samples from $0.30$ to $\SI{1.24}{\micro\kelvin}$.  
These temperatures are low enough to avoid contamination from nearby higher-partial-wave FFRs~\cite{Lecomte2024} while remaining high enough to ensure non-degenerate samples.
We fit the data, treating $R^\star$ and $\Gb$ as adjustable parameters, with a Floquet-based numerical approach that solves Eq.~\eqref{eqForphi3} with periodic $E_0(t)$ and deduces $L_2$ from Eqs.~\eqref{eq:L2_vs_mathcal_L2} and \eqref{eqL2} \cite{Sykes2017, SM}.
The model reproduces the measurements across the entire temperature range; see Fig.~\ref{L2vsT_pole_comparison}(a). From the fits we extract the on-resonance decay rate of the closed-channel molecule $\Gb^\mathrm{on}$
and the range parameter $R^\star$, which are both nearly temperature independent and given by~\cite{SM}
\be
\Gb^\mathrm{on}/(2\pi)=\SI{123(38)}{\kilo\hertz} \ \ {\rm and} \ \  R^\star = 10.0(2.3)\,\lvdW,
\label{values_Gb_Rs}
\ee
where $\lvdW = \SI{4.12}{\nano\meter}$ is the van der Waals length \cite{Maier2015}. 
As an independent check we perform bound-state spectroscopy, which yields $\Gb^\mathrm{on}/2\pi = \SI{112(30)}{\kilo\hertz}$ and $R^\star = 10.8(2.1)\, \lvdW$, consistent with Eq.~(\ref{values_Gb_Rs}), as well as the values of the differential magnetic moment $\delta\mu = {4.15 (18)}{\, \mu_B}$ and the FFR position $B_\textrm{res}$, such that $\Re (E_0) = \delta\mu(B-B_\textrm{res})$ (see End Matter).

We show in Fig.\ref{L2vsT_pole_comparison}(d) the two-body contact density $\mathcal{C}_2^{{\rm stat}} / n^2$, deduced from $\bar{L}_2$ using Eq.~\eqref{eqL2},  as a function of temperature. The measurements are well reproduced by~Eq.~\eqref{eqC2narrowFFR} (dashed blue line), evaluated with the parameters from Eq.~(\ref{values_Gb_Rs}); the shaded band reflects the uncertainties in these parameters.  These findings demonstrate a high degree of control over the observed two-body losses and establish their direct connection to the two-body contact density, whose dynamical evolution we resolve \cite{SM}.




  \begin{figure}[t!]
    \centering
	\includegraphics[width=\columnwidth]{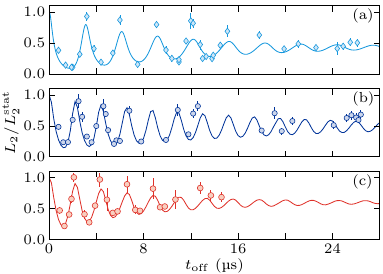}
	\caption{Coherent oscillations of the two-body loss rate.
Normalized loss rate $L_2/L_2^{{\rm stat}}$ as a function of $t_\mathrm{off}$ for a fixed $t_\mathrm{on} = \SI{3}{\micro\second}$.
The solid line corresponds to numerical solutions of Eq.~\eqref{eqForphi3} using the parameters from Eq.~\eqref{values_Gb_Rs}. 
The values of $(T, \Re(E_0^{\rm off})/h)$ are
$(\SI{0.41}{\micro\kelvin} , -\SI{436}{\kilo\hertz})$ for (a),
$(\SI{0.41}{\micro\kelvin} , -\SI{560}{\kilo\hertz})$ for (b), and
$(\SI{1.24}{\micro\kelvin} , -\SI{560}{\kilo\hertz})$ for (c).
}
    \label{interferences}
\end{figure}

Finally, we study the dependence of $L_2(t_\mathrm{on}, t_\mathrm{off})$ on $t_{\rm off}$ at fixed $t_\mathrm{on} = \SI{3}{\micro\second}$. The data, shown in Fig.~\ref{interferences}, reveal oscillations of the two-body loss rate, normalized by the asymptotic value $L_2^{{\rm stat}}$. We attribute these oscillations to the fact that, between pulses, the system is quenched out of resonance to the negative ${\Re}(E_0^{\rm off})$ side, where the closed-channel molecular state lies far below the open-channel threshold and is thus protected from dissociation. During this period, the closed-channel amplitude acquires a phase factor $\approx e^{-i E_0^{\rm off} t_{\rm off}/\hbar}$, interfering with the open-channel component which accumulates a phase $\approx e^{-i \hbar k^2 t_{\rm off} / m}$. The loss of contrast arises from thermal averaging over $k$ and from the finite closed-channel decay rate $\Gb^{\rm off}$, which remains relatively small in our case (see End Matter).

Similar oscillations have been observed in double-pulse experiments~\cite{Donley2002,Claussen2003,Elbaz2023}. A distinguishing feature of our experiment is the use of a periodic sequence of pulses. As shown in Fig.~\ref{interferences}(a), the resulting oscillations deviate markedly from a simple sinusoid, exhibiting an extended minimum followed by a sharp, asymmetric peak. This waveform reflects interference between multiple pathways for molecule formation, each associated with one of the $N_\mathrm{cycle}$ on-resonant pulses.

To gain a more quantitative insight on these oscillations we numerically solve Eq.~\eqref{eqForphi3}. The theory curve shown in Fig.\ref{interferences}(a) fits the data for a detuning $\Re(E_0^{{\rm off}})/h = \SI{-436 (7)}{\kilo\hertz}$, in good agreement with the expected detuning at the field $B_1$, $\Re(E_0^{\rm off})/h = \delta\mu \, (B_1-B_0)= \SI{-470 (40)}{\kilo\hertz}$ for the parameters determined in the End Matter. We then increase the SLS intensity by 40\%, shifting the  resonance pole to a lower magnetic field, $B_2 < B_1$, such that the off-resonant dimer energy is $\approx-h \times \SI{560}{\kilo\hertz}$. As a result, we observe faster oscillations, as shown in Fig.~\ref{interferences}(b), confirming that the oscillation frequency is set by the off-resonant dimer energy.

As we have mentioned, the reduction of contrast with increasing $t_{\rm off}$ is attributed to the finite temperature of the gas and to the intrinsic loss rate of the off-resonant molecular state, quantified by $\Gb^\mathrm{off}$ (see End Matter). To check this, we repeat the measurement at a higher temperature, $T = \SI{1.24}{\micro\kelvin}$. As shown in Fig.~\ref{interferences}(c), the contrast decays more rapidly, as predicted by the numerical model, confirming that it is the thermal decoherence that plays the key role in the observed damping.



\

In conclusion, we achieve a direct measurement of the real-time buildup of two-body correlations in a thermal Bose gas near a narrow resonance, enabled by submicrosecond interaction quenches. By connecting the temporal evolution of the two-body loss rate to Tan’s contact, we establish controlled dissipation as a powerful probe of short-range correlations, extending earlier studies of strongly interacting gases at equilibrium into the dynamical regime. The excellent agreement with a two-channel model demonstrates its predictive capacity and paves the way for exploring new quench protocols, pulse sequences, and interaction regimes. Looking ahead, the experimental and theoretical advances demonstrated here open promising routes to probe the nonequilibrium dynamics of Tan’s contact in degenerate Bose gases, including the most challenging strongly correlated regime.




\vskip10pt
\begin{acknowledgements}
  \textit{Acknowledgments: } We acknowledge fruitful discussions with Z. Hadzibabic and the members of the Bose-Einstein condensate team at LKB. This research was funded, in part, by Agence Nationale de la Recherche (ANR), projects ANR-20-CE30-0024, ANR-24-CE30-7961 and ANR-21-CE30-0033. For the purpose of open access, the author has applied a CC-BY public copyright licence to any Author Accepted Manuscript (AAM) version arising from this submission. This work was also funded by Region Ile-de-France in the framework of DIM QuanTiP, and supported by a grant from the ``Fondation CFM pour la Recherche''.
\end{acknowledgements}

\vspace{1cm}

\textit{Data availability}: The data that support the findings of this article are openly available  \cite{DataRepository}.

\bibliography{bib-final.bib}


\section*{End Matter}

\section{Determination of $\Gb$ and $R^\star$ from spectroscopy of the dimer energy}

  \begin{figure}[h!]
    \centering
	\includegraphics{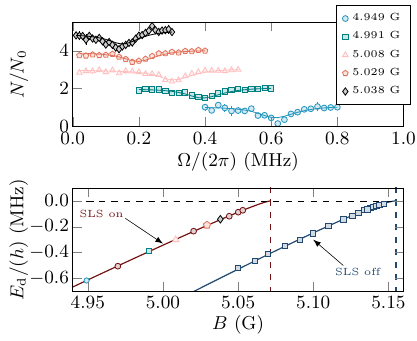}
	\caption{Determination of the narrow Fano-Feshbach resonance properties.
Top: Resonance spectra of atom losses as a function of the modulation frequency $\Omega$ for various magnetic fields (see legend). For visual clarity, the different curves are vertically offset by consecutive integer values. The solid lines are fits to the data using Eq.~\eqref{eqForphi3}.
Bottom: Resonance frequency as a function of magnetic field, for $I_0=0$ (SLS off) and $I_0 = 15 \mu\rm{W/\mu m^2}$ (SLS on). The solid line shows the expected dressed Feshbach dimer energy dependence on magnetic field with the parameters listed in Table~\ref{tab:FFR}.}
    \label{RFresonance}
\end{figure}

To determine the closed-channel dimer energy as a function of magnetic field, we apply the SLS beam with a temporal intensity profile $I(t) = I_0 + I_1 \sin^2(\Omega t / 2)$. For half of the data we use $I_0=0$ and $I_1=1.0~\mu\mathrm{W}/\mu\mathrm{m}^2$ (SLS off in Fig.\ref{RFresonance}), which places the FFR pole at $B_0 \approx \SI{5.15}{\gauss}$. For the other half we set $I_0=15\mu\mathrm{W}/\mu\mathrm{m}^2$ and $I_1=1.5~\mu\mathrm{W}/\mu\mathrm{m}^2$ (SLS on in Fig.~\ref{RFresonance}), shifting the pole to $B_1 \approx \SI{5.07}{\gauss}$. In both cases the average intensity is $I_0 + I_1/2$.
We then scan the frequency $\Omega$ around the molecular binding energy and observe resonant atom loss when $\hbar \Omega$ matches the real part of the closed-channel binding energy, signaling the formation of weakly bound molecules that subsequently decay (see Fig.~\ref{RFresonance} top panel). The extracted binding energies as a function of magnetic field are shown in Fig.~\ref{RFresonance} (bottom panel).

The only differences between the two datasets are the effective resonance position, determined by the SLS beam, and the inverse lifetime of the molecular state, $\Gb$, which increases with SLS intensity. 
The value of $\Gb^{\rm off}$ is much smaller than $\Gb^{\rm on}$, yet remains finite. Its value is likely set by photodissociation from the infrared trapping lasers. We observed that $\Gb^{\rm off}$ remained unchanged for $I_1$ up to $2.0~\mu\mathrm{W}/\mu\mathrm{m}^2$ (a factor of two increase), indicating that the contribution from the residual SLS average intensity, $I_1/2$, is negligible in this regime.

The best-fit parameters characterizing the resonance spectra are reported in Tab.~\ref{tab:FFR}. The line shown in Fig.~\ref{RFresonance} corresponds to the following expression for the real part of the dressed Feshbach dimer energy~{\cite{Zhou2021}
\be 
\Re(E_{{d}}) = - \frac{\hbar^2}{4 m {R^\star}^2} \left( \sqrt{1 + 4R^\star/a'} - 1 \right)^2 \, ,
\label{eqbound} 
\ee 
with
\be 
a' = -\frac{\hbar^2}{m R^\star \delta \mu (B-B_\mathrm{res})}\, ,
\label{eqscatt}
 \ee
where $B_\mathrm{res}$ is the position of the FFR; $B_\mathrm{res}= B_0$ (resp. $B_\mathrm{res}= B_1$) when the SLS is off (resp. on).

\renewcommand{\arraystretch}{1.5} 

\begin{table}[h]
 \caption{Values of the FFR position, $B_\mathrm{res}$, differential magnetic moment, $\delta\mu$, and $R^\star$ in the presence and absence of the spin-dependent light shift.}
 \label{tab:FFR}
\begin{ruledtabular}
\begin{tabular}{ccc}
 & $I_0 = 0$ (SLS off) & $I_0 \neq 0$ (SLS on)\\
\hline
$\delta \mu$ ($\mu_\text{B}$)& 4.35\,(16) & 4.15\,(18) \\
$B_\mathrm{res}$ (G) & 5.154\,(2) & 5.073 \,(2) \\
$R^\star$ (\lvdW) & 10.2\,(1.8)& 10.8\,(2.1)\\
$\Gb/(2\pi)$ (kHz) & 20 (10) & 112\,(30)\\
\end{tabular}\label{tab1}
\end{ruledtabular}
\end{table}




\newpage

\section*{Supplemental Material\\Two-Body Contact Dynamics in a Bose Gas near a Fano-Feshbach Resonance}

\renewcommand{\thefigure}{S\arabic{figure}}
\setcounter{figure}{0} 
\renewcommand{\theequation}{S\arabic{equation}}
\setcounter{equation}{0} 

\appendix

\twocolumngrid 

\newcommand{\1}{1}
\newcommand{\2}{2}

\section{Density-dependence of losses}

\begin{figure}[ht!]
\centering
\includegraphics[width=\columnwidth]{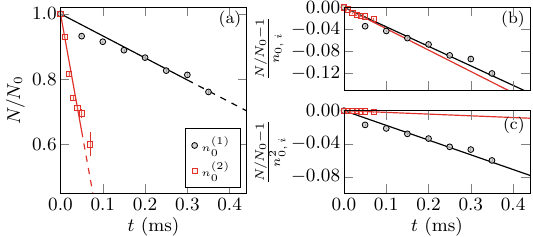}
\caption{Loss rate as a function of density.
(a) Atom number evolution as a function of time $t$ for two different initial densities (see legend and text), following a quench to the FFR. At short times, losses exhibit a linear dependence on time.
(b) Atom number normalized by the initial peak density. The data collapse onto a single linear curve, consistent with a two-body loss process.
(c) Atom number normalized by the square of the initial peak density. The differing evolutions clearly indicate that the losses are incompatible with a three-body process.}
\label{lossesdensity}
\end{figure}

In this section, we describe the procedure that leads to the conclusion that the losses in our experiment have a two-body nature.
The differential equation
\be
\dot n = -\mathcal{L}_\alpha n^\alpha
\ee
describes the evolution of the density in the presence of a loss mechanism involving $\alpha$-body collisions. 
In our experiment, the single-particle lifetime is on the order of \SI{15}{\second}, far exceeding the timescales probed here. The dominant loss mechanisms are therefore either dimer losses involving photon-assisted photodissociation ($\alpha=2$) or intrinsic three-body losses ($\alpha=3$). Assuming a Gaussian spatial distribution, and integrating over space, the total atom number then evolves according to:
\be
\dot N = - \frac{\mathcal{L}_\alpha n_0^{\alpha-1}}{\alpha^{3/2}} N \, .
\label{eqdiff_N_alpha}
\ee
Its solution, neglecting the time-dependence of $\mathcal{L}_\alpha$\,, is given by
\be
N(t) = \frac{N_0}{\left[1 + \left( \frac{\alpha - 1}{\alpha^{3/2}} \right) \mathcal{L}_\alpha n_{0,i}^{\alpha - 1} t \right]^{1/(\alpha - 1)}}
\label{N(t)}
\ee
where $n_{0,i} = N_0/V$ is the peak density at $t=0$, $N_0$ the initial atom number and $V = \pr{2\pi \kb T \big/ \pc{m \bar \omega^2}}^{3/2}$ the volume.  
We restrict to short enough times such that the temperature of the cloud changes by less than 15\% of its initial value. For these short times, Eq.~\eqref{N(t)} simplifies to
\be
\frac{N(t)}{N_0} \approx 1 - \mathcal{L}_\alpha \frac{n_{0,i}^{\alpha - 1}}{\alpha^{3/2}} t \, .
\label{eqNshortT}
\ee

We investigate two thermal samples at the same temperature, $T = \SI{0.9}{\micro\kelvin}$, but with different geometric mean trap frequencies: $\bar{\omega}^{(1)} = 2\pi \times \SI{141}{\hertz}$ and $\bar{\omega}^{(2)} = 2\pi \times \SI{212}{\hertz}$. The initial atom numbers are such that $N_0^{(2)} \approx 3 \, N_0^{(1)}$, and therefore $n_{0,i}^{(2)} \approx 10 \, n_{0,i}^{(1)}$.

In Fig.~\ref{lossesdensity}(a), we show the evolution of $N(t)/N_0$ as a function of $t$, confirming the initial linear decrease in atom number for both cases. We then examine the power-law dependence of the loss rate on the density. 
From Eq.~\eqref{eqNshortT} we expect that the quantity
$[N(t)/N_0-1] / n_{0,i}^{\alpha - 1} \approx  - (\mathcal{L}_\alpha / \alpha^{3/2})\, t$ is independent of $n_{0,i}$.  We find that the experimentally measured values for this quantity do indeed approximately collapse on a single curve if we assume $\alpha=2$ (Fig.~\ref{lossesdensity}b), but not for $\alpha=3$ (Fig.~\ref{lossesdensity}c).
We conclude that the losses are predominantly two-body in nature. 
We performed the same analysis for different temperatures and densities, and reached the same conclusion.

To theoretically explain why three-body losses are negligible in our regime, one would need to solve the three-body problem for lossy and narrow two-body resonances (e.g., by generalizing the method developed in Ref.~\cite{Rem2013}), which is beyond the scope of this work. We can nevertheless provide a simple order-of-magnitude estimate in the large-$R^\star$ limit, where  the relaxation rate of bare closed-channel molecules colliding with free atoms, $\Gamma_{\rm 3B}$, is on the order of $\hbar \lvdW n / m$. For the typical densities used in this work, this yields $\Gamma_{\rm 3B} \approx  \SI{0.016}{\per\milli\second}$, which is  more than 4 orders of magnitude smaller than the photo-dissociation loss rate $\Gb^{\rm on} \approx \SI{754}{\per\milli\second}$ from Eq.~(7).

\section{Cycling procedure}
As described in the main text, we quench the system to resonance for a duration $t_\mathrm{on}$, followed by an off-resonant period $t_\mathrm{off}$. The total exposure time at resonance is given by $t_\mathrm{exp} = N_\mathrm{cycles} \, t_\mathrm{on}$. We use integer values of $N_\mathrm{cycles}$, corresponding to square pulses generated by a waveform generator that modulates the radiofrequency amplitude driving an acousto-optic modulator.
Although $t_\mathrm{exp}$ takes discrete values,  the variation of $N$ over one cycle is small, and $N$ can be considered to be a smooth differentiable function of $t_\mathrm{exp}$.
In practice, the maximum number of cycles ranges from 10 to 5000 depending on the chosen value for $t_\mathrm{on}$.

\section{Extracting $L_2$ in a Thermal Gas with Evolving Temperature}
\label{secTemp}

\begin{figure}[ht!]
\centering
\includegraphics[width=\columnwidth]{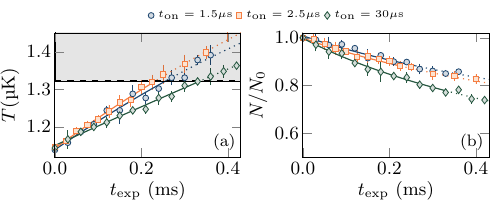}
\caption{Temperature and atom number evolution. (a) Temperature evolution as a function of time after quenching a thermal sample with initial temperature $T = \SI{1.15}{\micro\kelvin}$ to the Fano–Feshbach resonance, for different interrogation times $t_\mathrm{on}$ (see legend) and $t_\mathrm{off} = \SI{26}{\micro\second}$. The shaded area corresponds to a temperature increase greater than 15\% of the initial temperature, where we exclude all data from our analysis. The lines are fits to the data using a polynomial function.
(b) Normalized atom number evolution as a function of time. The fitted curves correspond to the resolution of Eq.~\eqref{NevolvTaccounted} accounting for the temperature evolution extracted from the fits shown in (a).}
\label{heating}
\end{figure}

Atom loss is accompanied by heating in the harmonic trap, as shown in Fig.~\ref{heating}(a), resulting in a variation of the effective volume due to the thermal expansion of the cloud.
To ensure accurate determination of the two-body loss rate, we restrict our analysis to the regime where the temperature increases by no more than 15\%, as indicated by the shaded area in Fig.~\ref{heating}(a). 

Equation~(1) of the main text yields the differential equation
\be
\label{NevolvTaccounted}
\frac{dN}{dt_\mathrm{exp}} \ = \ - L_2 (\bar T) \,N(t_{\mathrm{exp}})^2 \,\big/{\pr{ \frac{4\pi k_B}{m \bar{\omega}^2}  \, T(t_{\mathrm{exp}})}^{3/2}} \, ,
\ee
which we solve numerically, with $T(t_{\mathrm{exp}})$ obtained by fitting the measured temperature evolution with a second order polynomial (see Fig.~\ref{heating}). This allows us to account for the dilatation of the sample over time. We use the fitted temperature curve to determine the time-average temperature $\bar T$ which is used in the computation of the corresponding two-body loss rate.

\section{Temperature dependence of the loss  resonance feature}
\label{secFFRvsT}

\begin{figure}[ht!]
\centering
\includegraphics[width=\columnwidth]{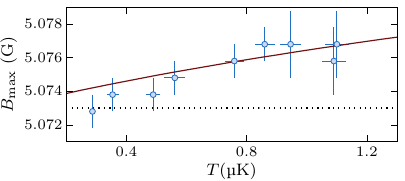}
\caption{
Magnetic field $B_{\rm max}$ at which the loss rate is maximal, shown versus the sample temperature. The trap frequencies are fixed at $(\omega_x,\omega_y,\omega_z) = 2\pi \times (58,358,292)$ Hz for all data. The dotted line indicates the FFR position obtained from closed-channel molecular spectroscopy. The solid line is the theoretical prediction for $B_{\rm max}$ which maximizes Eq.~(5) of the main text.
}
\label{peaklossvstemp}
\end{figure}

A consequence of the narrowness of the FFR is that for nonzero temperature, the maximum loss rate does not occur exactly when the scattering length diverges. 
For a fixed trap frequency, we show in Fig.~\ref{peaklossvstemp} that the magnetic field at which losses are maximal, $B_\mathrm{max}$, indeed shifts with temperature.
We note for completeness that $t_\textrm{exp}$ was adjusted at each temperature such that the maximum loss corresponds to 30\% of the initial atom number.

The experimentally determined  $B_\mathrm{max}$ is in good agreement with the theoretical prediction (see solid line in Fig.~\ref{peaklossvstemp}) obtained from Eq.~(5)  without adjustable parameters (we use $R^\star$ and $\Gb^\mathrm{on}$ from Eq.~(7), and $\delta \mu$ and $B_\mathrm{res}$ from Tab.~I, right column).

\section{Fitting procedure to extract $R^\star$ and $\Gb^{\mathrm{on}}$}
\label{secFittingprocedure}

  \begin{figure}[ht!]
    \centering
	\includegraphics[width=\columnwidth]{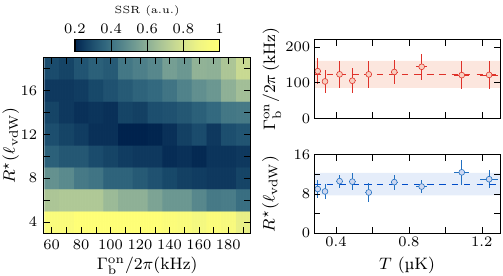}
	\caption{
     Temperature-dependence of fitted parameters. Left panel: Sum of squared residuals (SSR) for the data at $T = \SI{1.09(4)}{\micro\kelvin}$. Right panels: 
     $\Gb^\mathrm{on}$ and $R^\star$ {\it vs.} temperature. Here dashed lines indicate temperature averages and shaded regions show the associated standard deviations.}
    \label{numericalfit}
\end{figure}

To compare the experimental data with the numerical simulations for $L_2(t_\textrm{on}, t_\textrm{off})$, we use a Floquet-based numerical approach that solves Eq.~(4) with periodic $E_0(t)$ and averages the instantaneous loss rate over a full modulation cycle, $t_\mathrm{on}+t_\mathrm{off}$ (see Ref.~\cite{Sykes2017} and the last section of this Supplemental Material), which depends on both on- and off-resonant parameters. 

For the off-resonant scattering length we use $a'_\mathrm{off} \approx 46\, a_0$ (see End Matter).
For the on-resonant scattering length $a'_\mathrm{on}$, we use the value that maximizes Eq.~(5) of the main text for each temperature (see previous section). 
Regarding the molecular decay rate $\Gb(t)$, we assume that $\Gb(t) = \Gb^\mathrm{on}$ during the on-resonant interval $t_\mathrm{on}$ and $\Gb(t) = \Gb^\mathrm{off}$ during the off-resonant time $t_\mathrm{off}$.
For each temperature, we fit the data by minimizing the sum of the squared residuals with respect to the numerical simulation, while varying $R^\star$, $\Gb^\mathrm{on}$, and $\Gb^\mathrm{off}$. 
We observed that $\Gb^\mathrm{off}$ had little impact on the fits, and therefore fixed it to the value $2\pi \times \SI{20}{\kilo\hertz}$ obtained by modulation spectroscopy (see Tab.~I).

The fitting procedure thus reduces to a two-dimensional optimization over $R^\star$ and $\Gb^\mathrm{on}$. Specifically, we vary $R^\star$ in the range $4\,\lvdW$ to $18\,\lvdW$, and $\Gb^\mathrm{on}$ from \SI{60}{\kilo\hertz} to \SI{190}{\kilo\hertz}. For all temperatures, we observe a well-defined optimal region in parameter space (see Fig.~\ref{numericalfit}, left panel). The error bars on the fitted parameters are obtained using a bootstrap method. In the right panels of Fig.~\ref{numericalfit}, we check the temperature independence of $R^\star$ and $\Gb^\mathrm{on}$, respectively. The dashed lines and shaded areas represent the temperature-averaged values and their standard deviations, yielding $R^\star = 10.0(2.3)\,\lvdW$ and $\Gb^\mathrm{on}/2\pi = \SI{123(38)}{\kilo\hertz}$.

\section{Evolution of $L_2$ as a function of $t_\mathrm{on}$ and $t_\mathrm{off}$}
\label{secL2vstoff}

  \begin{figure}[t!]
    \centering
	\includegraphics[width=\columnwidth]{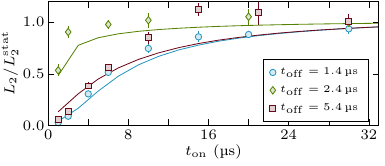}
	\caption{Evolution of $L_2$ as a function of $t_\mathrm{off}$ and $t_\mathrm{on}$.
Buildup of the two-body loss rate $L_2(t_\mathrm{on})$ for various out-of-resonance duration $t_\mathrm{off}$ (see legend) at a fixed temperature of $T = \SI{0.41}{\micro\kelvin}$. Solid lines are numerical solutions of Eq.~(4) using $\Gb/2\pi = \SI{123}{\kilo\hertz}$ and $\Re(E_0)/h = -\SI{560}{\kilo\hertz}$.}
    \label{interferencesSM}
\end{figure}

We show in Fig.~\ref{interferencesSM} the buildup of the two-body loss rate, normalized by its asymptotic value $L_2^{\rm stat} = \SI{283(19)}{\micro\meter^3/\second}$, as a function of $t_\mathrm{on}$ for three different values of $t_\mathrm{off}$ (see legend).
Although the overall behavior of $L_2$ is non-monotonic with respect to $t_\mathrm{off}$ (see Fig.~4 and its discussion in the main text), we observe a monotonic increase of $L_2$ with $t_\mathrm{on}$ for each individual $t_\mathrm{off}$.
Importantly, we find that for a fixed modulation period, i.e., the same value of $t_\mathrm{on} + t_\mathrm{off}$, the measured two-body loss rate can vary significantly.
For example, for $t_\mathrm{off} = \SI{1.4}{\micro\second}$ and $t_\mathrm{on} = \SI{2}{\micro\second}$, we find $L_2 / L_2^{\rm stat} = 0.089(7)$, whereas for the same total modulation period but with $t_\mathrm{off} = \SI{2.4}{\micro\second}$ and $t_\mathrm{on} = \SI{1}{\micro\second}$, we measure $L_2 / L_2^{\rm stat} = 0.53(6)$.\\~\\

\section{Derivation of the evolution equations \\for the closed-channel amplitude}

We solve the problem of two atoms in a large volume, which can be taken to be a unit volume to alleviate notations. This is equivalent to the second order nonequilibrium virial expansion of Refs.~\cite{Sun2020,Cui2024,Yang2023}. Although we deal with identical bosons, for future reference we provide the derivation of the evolution equations for two distinguishable atoms of generally different masses. Consider the two-channel Hamiltonian (we set $\hbar=1$) 
\begin{widetext}
\begin{align}\label{SM:Ham}
&\hat{H}=\int d^3r \sum_{\sigma=\1,\2} 
\hat{\psi}^\dagger_\sigma({\bf r}) 
\left(-\frac{\nabla^2_{\bf r}}{2m_\sigma}\right) 
\hat{\psi}_\sigma({\bf r}) 
+
\hat{\psi}^\dagger_b({\bf r}) 
\left[-\frac{\nabla^2_{\bf r}}{2(m_\1+m_\2)}+\nu(t)\right] 
\hat{\psi}_b({\bf r}) 
\nonumber\\
&-\eta\int d^3r \ d^3y \ \delta_{r_0}({\bf y})\,\left[\hat{\psi}_\1^\dagger({\bf r}+\mu{\bf y}/m_\1) \, \hat{\psi}_\2^\dagger ({\bf r}-\mu {\bf y}/m_\2) \, \hat{\psi}_b({\bf r})+h.c.\right],
\end{align}  
\end{widetext}
where $\hat{\psi}^\dagger_\1({\bf r})$ and $\hat{\psi}^\dagger_\2({\bf r})$ create atoms of masses $m_\1$ and $m_\2$, respectively,  
$\hat{\psi}^\dagger_b({\bf r})$ 
is the creation operator of a bare closed-channel molecule, and $\mu=m_\1 m_\2/(m_\1+m_\2)$ is the effective mass. To regularize the model we use the delta-shell representation with $\delta_{r_0}({\bf y})=\delta(|{\bf y}|-r_0)/(4\pi r_0^2)$, were $r_0$ is assumed to be the smallest length scale in the problem. As we will see, the model (\ref{SM:Ham}) has a well-defined zero-range limit such that $r_0$ drops out of the final equations and the bare parameters $\nu$ (complex) and $\eta$ (real) are expressed in terms of the renormalized physically meaningful quantities: the complex detuning $E_0(t)=\Re\pr{E_0(t)}-i\,\Gb(t)/2$ and the real range parameter $R^\star$ characterizing the resonance width (a standard procedure also used, {e.g.}, in Refs.~\cite{Pricoupenko2013,Zhou2021}). We note that it is possible to add a direct  interaction between atoms in the open channel, characterized in the zero-range limit by the background scattering length (see, {e.g.}, Refs.~\cite{Braaten2008PRA,Yang2023,Pricoupenko2019}); we have checked that for our experimental parameter regime, this does not lead to any significant changes of the results, and we omit the background interaction to simplify the discussion.

The center-of-mass degree of freedom separates and the relative two-body wave function can in general be written as 
\begin{align}
&\hspace{-0.5cm}\int d^3r \ d^3y \ \Psi({\bf y},t) \ \hat{\psi}^\dagger_\1 ({\bf r}+\mu {\bf y}/m_\1 ) \, \hat{\psi}^\dagger_\2 ({\bf r}-\mu {\bf y}/m_\2 )\ket{0}\nonumber\\
&\hspace{4cm}+\int d^3 r \ \phi(t) 
\, \hat{\psi}^\dagger_b({\bf r}) 
\ket{0},\label{SM:TwoBodyState}
\end{align}
where $\ket{0}$ is the vacuum state. The evolution of the amplitudes $\Psi$ and $\phi$ is governed by the coupled Schr\"odinger equations
\begin{equation}
i\partial_t\Psi({\bf y},t)= \,- \, \frac{\nabla^2_{\bf y}}{2\mu} \ \Psi({\bf y},t)-	\eta\,\delta_{r_0}({\bf y})\,\phi(t)
\label{TwoBodyPsiInt}
\end{equation}
and
\begin{equation}
i\partial_t\phi(t)=\nu(t)\,\phi(t)-	\eta\int d^3 y \ \delta_{r_0}({\bf y})\,\Psi({\bf y},t).
\label{TwoBodyphiInt}
\end{equation}

Equation~(\ref{TwoBodyPsiInt}) describes free motion everywhere in space except for the surface of the sphere with radius $r_0$. It is useful to rewrite this equation in the integral form
\begin{multline}
\Psi({\bf y},t) \ = \ \Psi_0({\bf y},t) 
\\ \ \ \ + \, i \, \eta \, \int_{-\infty}^t 
dt' 
\int 
d^3y' \  G({\bf y}-{\bf y}'\!,t-t') \ \delta_{r_0}({\bf y}') \ \phi(t'),
\label{TwoBodyPsiInt1}
\end{multline}
where the Green function 
\begin{equation}\label{G}
G({\bf y},t)=\frac{e^{iy^2\mu/2t}}{(2\pi i t/\mu)^{3/2}}
\end{equation}
solves
\begin{equation}\label{EqforG}
[i\partial_t+\nabla^2_{\bf y}/(2\mu)]G(y,t)=i \,\delta(t)\,\delta({\bf y})
\end{equation}
and $\Psi_0({\bf y},t)$ is a solution of Eq.~(\ref{TwoBodyPsiInt}) with $\eta$ set to zero. For our scattering problem this is the incoming plane wave $\Psi_0({\bf y},t)=e^{i{\bf k}\cdot{\bf y}-ik^2t/(2\mu)}$ with momentum ${\bf k}$. 

The idea of passing from Eq.~(\ref{TwoBodyPsiInt}) to Eq.~(\ref{TwoBodyPsiInt1}) is that, if we now set $|{\bf y}|=r_0$, Eqs.~(\ref{TwoBodyphiInt}) and (\ref{TwoBodyPsiInt1}) form a closed system of equations for $\phi(t)$ and $\Psi({\bf y},t)$, where ${\bf y}$ is on the sphere. Moreover, Eq.~(\ref{TwoBodyPsiInt1}) [as well as the original Hamiltonian (\ref{SM:Ham})] conserves angular momentum and allows for a separate description of each partial wave. In our case the $s$-wave scattering is dominant and for brevity we concentrate on this channel, which means that we deal with two time-dependent functions $\phi(t)$ and $\Psi(r_0,t)$. Working out the integral over $d^3 y$ in the limit of small $r_0$  Eq.~(\ref{TwoBodyPsiInt1}) reduces to 
\begin{equation}
\Psi(r_0,t)=e^{-ik^2t/(2\mu)}+\frac{\eta\phi(t)}{2\pi r_0/\mu} + i\eta \int_{-\infty}^t \frac{\phi(t')-\phi(t)}{[2\pi i (t-t')/\mu]^{3/2}}dt'
\label{TwoBodyPsiInt2BG}
\end{equation}
where, on the right-hand side we have neglected terms of order $\mu^2 \eta r_0^2|\partial_t \phi(t)|$. They are smaller than the integral term by a factor $\sqrt{\mu r_0^2/\tau}$ where $\tau$ is the characteristic time scale of variations of $\phi$. For completeness we also explicitly write Eq.~(\ref{TwoBodyphiInt}) projected on the $s$-wave channel
\begin{equation}
i\partial_t\phi(t)=\nu(t)\,\phi(t)-	\eta \,\Psi(r_0,t).
\label{TwoBodyphiIntFin}
\end{equation}
The function $\Psi(r_0,t)$ is eliminated from Eqs.~(\ref{TwoBodyPsiInt2BG}-\ref{TwoBodyphiIntFin}) and after partial integration the resulting equation for the closed-channel amplitude reads
\begin{align}
&i\partial_t\phi(t)-[\nu(t)-\mu\,\eta^2/(2\pi r_0)]\phi(t)\nonumber\\
&-\frac{\eta^2}{\sqrt{2i}}\left(\frac{\mu}{\pi}\right)^{3/2}\int_{-\infty}^{t}\frac{\partial_{t'}\phi(t')}{\sqrt{t-t'}}dt'=-\eta e^{-ik^2t/(2\mu)}.\label{TwoBodyphiIntFin2}
\end{align}

In the stationary case (time-independent $\nu$) Eq.~(\ref{TwoBodyphiIntFin2}) is solved by $\phi(t)=\phi_0 e^{-ik^2t/(2\mu)}$ with 
\begin{equation}
\phi_0=-\frac{\eta}{k^2/(2\mu)-[\nu-\mu\eta^2/(2\pi r_0)]+i\mu\eta^2 k/(2\pi)}\label{phi0}
\end{equation}
and Eq.~(\ref{TwoBodyPsiInt1}) becomes
\begin{equation}
\Psi({\bf y},t)=e^{i{\bf k}\cdot{\bf y}-ik^2t/(2\mu)}+\frac{\mu\eta \phi_0}{2\pi}\frac{e^{iky-ik^2t/(2\mu)}}{y}
\label{TwoBodyPsiStat}
\end{equation}
valid for $y\geq r_0$. In Eq.~(\ref{TwoBodyPsiStat}) one recognizes the scattering amplitude $f(k)=\mu\eta \phi_0/(2\pi)$. Identifying
\begin{equation}\label{alphavsRs}
\eta=\sqrt{\frac{\pi}{\mu^2 R^\star}}
\end{equation}
and
\begin{equation}\label{TildeE0vsE0}
E_0=\nu-\frac{1}{2\mu R^\star r_0}
\end{equation}
we recover the standard narrow-resonance structure of the scattering amplitude
\begin{equation}\label{fscat}
f(k)=-\frac{1}{R^\star (k^2-2\mu E_0)+ik}
\end{equation}
and the stationary closed-channel amplitude equals
\begin{equation}\label{Statphi}
\phi_0=-\frac{2\sqrt{\pi/R^\star}}{k^2-2\mu E_0+ik/R^\star}.
\end{equation}
Substituting Eqs.~(\ref{alphavsRs}-\ref{TildeE0vsE0}) into Eq.~(\ref{TwoBodyphiIntFin2}) we obtain the evolution equation with renormalized parameters
\begin{align}
&\hspace{-0.7cm}i\partial_t\phi(t)-E_0(t)\,\phi(t)-\frac{1}{\sqrt{2\pi i\mu}\,R^\star}\int_{-\infty}^{t}\frac{\partial_{t'}\phi(t')}{\sqrt{t-t'}}dt'\nonumber\\
&\hspace{3cm}=-\,\sqrt{\frac{\pi}{\mu^2 R^\star}} \ e^{-ik^2t/(2\mu)}.\label{TwoBodyphiIntFin3}
\end{align}
To reproduce Eq.~(4) of the main text we should set $\mu=m/2$ and multiply the right-hand side of Eq.~(\ref{TwoBodyphiIntFin3}) by $\sqrt{2}$, which takes into account the correct symmetrization of the incoming wave for identical bosons $\Psi_0({\bf y},t)=\sqrt{2}\cos({\bf k}{\cdot}{\bf y})e^{-ik^2t/(2\mu)}$. Accordingly, the stationary contact density of a thermal Bose gas reported in Eq.~(5) of the main text is obtained from Eq.~(\ref{Statphi}) multiplied by $\sqrt{2}$ and where we set $\mu=m/2$.

We emphasize that our theory is valid in the zeroth order in the small parameter $\sqrt{\mu r_0^2/\tau}$ [see discussion after Eq.~(\ref{TwoBodyPsiInt2BG})]. For the stationary case the characteristic time of variation of $\phi$ is $\tau\sim \mu/k^2$ and the small parameter becomes $r_0k$. In practice this means that we cannot go beyond ultracold temperatures and describe quenches on timescales associated with the interaction range. 

As we explain in the main text, the knowledge of $\phi$ for a pair of atoms in a unit volume can be used for determining the contact density. This statement is based on the observation, which holds also in the nonstationary case under the condition $\sqrt{\mu r_0 /\tau}\ll 1$, that the closed-channel amplitude $\phi(t)$ is related to the $1/y$ singularity of $\Psi(y,t)$ by $\phi(t)=\sqrt{4\pi R^\star}\lim_{y\rightarrow 0}y\,\Psi(y,t)$. In the stationary case this follows directly from Eq.~(\ref{TwoBodyPsiStat});
in the general case it can be seen by keeping in Eq.~(\ref{TwoBodyPsiInt2BG}) only terms diverging as $1/r_0$. Namely, we have $\Psi(r_0,t)=\mu\eta\phi(t)/(2\pi r_0)$, which, by using Eq.~(\ref{alphavsRs}), reduces to $\phi(t)=\sqrt{4\pi R^\star}\,r_0\,\Psi(r_0,t)$.

Finally we note that in the broad-resonance limit ($R^\star{\to}\,0$),  the $\dot{\phi}$ term drops out in Eq.~(4) of the main text, which implies that the function $\phi(t)/\sqrt{R^\star} \propto \lim_{r\to0} r\,\Psi(\rr,t)$ solves the integral equation already obtained in Ref.~\cite{Qi2021}.


\section{Single-quench evolution of $\mathcal{C}_2(t)$}

  \begin{figure}[ht!]
    \centering
	\includegraphics[width=\columnwidth]{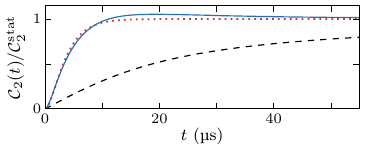}
	\caption{Single-quench evolution of $\mathcal{C}_2 (t)$. Evolution of $\mathcal{C}_2 (t)$, scaled by its stationary value, for $T=\SI{0.4}{\micro\kelvin}$. Blue line: numerical computation for $R^\star = 10 \lvdW$ and $\Gb / 2\pi = \SI{123}{\kilo\hertz}$. Red dotted line: asymptotic expression in the lossy narrow resonance limit, see main text, for $\Gb / 2\pi = \SI{123}{\kilo\hertz}$. Black dashed line: asymptotic expression in the lossless broad resonance limit, see main text.}
    \label{figNumericComparison}
\end{figure}

Based on Eq.~(6) of the main text, we take the inverse Laplace transform of $\phi(s)$ and numerically compute the thermal average of $|\phi(t)|^2$, which allows us to extract the single-quench evolution of $\mathcal{C}_2$ for our experimental parameters $R^\star = 10 \lvdW$ and $\Gb/2\pi = \SI{123}{\kilo\hertz}$ (see Eq.~(7) of main text). In Fig.~\ref{figNumericComparison}, we compare the numerical results, for $T=\SI{0.4}{\micro\kelvin}$, to the asymptotic formulas obtained in the lossless broad limit, $\mathcal{C}_2(t)/\mathcal{C}_2^{\rm stat}~=2\ {\rm arctan}({t/\tau_T})/\pi$, and lossy narrow resonance limit, $\mathcal{C}_2(t)/\mathcal{C}_2^{\rm stat}~=(1 - e^{-\Gb t/2})^2$. We note that for narrow resonances, $\mathcal{C}_2(t)$ can become non-monotonic, whereas for broad resonances this only occurs when quenching to the $\Re(E_0)<0$ side of the resonance~\cite{Cui2024}.

\section{Floquet analysis}

For time-periodic $E_0(t)$ with period $2\pi/\Omega$ we use the formalism of Ref.~\cite{Sykes2017} and solve the integro-differential Eq.~(\ref{TwoBodyphiIntFin3}) by decomposing $\phi(t)$ in the Floquet channels
\begin{equation}
\phi(t)=\sum_{n=-\infty}^\infty \phi_n e^{-i\omega_n t},\label{ExpphiBG}
\end{equation}
where $\omega_n=\Omega n+k^2/(2\mu)$. Substituting this expansion into Eq.~(\ref{TwoBodyphiIntFin3}) and using the equality
\begin{equation}\label{equality}
\int_{-\infty}^t \frac{e^{-i\omega t'}-e^{-i\omega t}}{(t-t')^{3/2}}dt'=-e^{-i\omega t}\sqrt{2\pi|\omega|}(1-i \ {\rm sign} \ \omega),
\end{equation}
valid for real $\omega$, we obtain
\begin{align}\label{phiFloquetFin}
\left[\frac{\pi}{2}\sqrt{\frac{|\omega_n|}{i\mu }}(1-i \ {\rm sign} \ \omega_n)-\omega_n\right]\phi_n&+\sum_{m=-\infty}^\infty D_{n-m}\phi_m\nonumber\\
&=\sqrt{\frac{\pi}{\mu^2 R^*}}\delta_{n0},
\end{align}
where $D_n$ is the Fourier transform
\begin{equation}\label{nuFourier}
D_n=\frac{\Omega}{2\pi}\int_0^{2\pi/\Omega}e^{i\Omega n t}E_0(t)dt
\end{equation}
and $\delta_{nm}$ is the Kronecker delta. Denoting by $E_0^{\rm on}$ the value of the (complex) detuning at times $0<t<t_{\rm on}$ and $E_0^{\rm off}$ at times $t_{\rm on}<t<t_{\rm on}+t_{\rm off}$ the Fourier transform of the infinite sequence of square pulses ($t_{\rm on}+t_{\rm off}=2\pi/\Omega$) reads
\begin{equation}\label{D0}
D_{n=0}=\frac{\Omega}{2\pi}(E_0^{\rm on}t_{\rm on}+E_0^{\rm off} t_{\rm off}),
\end{equation}
\begin{equation}\label{Dn}
D_{n\neq 0}=\frac{e^{i\Omega n t_{\rm on}}-1}{2\pi i n}(E_0^{\rm on}-E_0^{\rm off}).
\end{equation}
We solve Eq.~(\ref{phiFloquetFin}) numerically by introducing a cutoff for sufficiently high $|n|$ ensuring convergence. The vector $\phi_n$ [or $\phi(t)$ given by Eq.~(\ref{ExpphiBG})] can then be used for calculating all relevant observables (the contact, the instantaneous or averaged loss rate, etc.)

\end{document}